\def\hone{\ifmmode{\rm H \/ {\sc i}}\else{H\/{\sc i}}\fi} 
\def\lesssim{\mathrel{\hbox{\rlap{\hbox{\lower4pt\hbox{$\sim$}}}\hbox{$<$}}}}
\def\gtrsim{\mathrel{\hbox{\rlap{\hbox{\lower4pt\hbox{$\sim$}}}\hbox{$>$}}}}

\documentclass[useAMS,usenatbib]{mn2e}
\usepackage{epsfig}
\begin{document}
\newcommand{\average}{\bar}
\newcommand{\science}{{ Science}}
\newcommand{\physrev}{{ Phys. Rev.}}
\newcommand{\pasp}{{ PASP}}
\newcommand{\prd}{{ Phys. Rev. D}}
\newcommand{\apj}{{ ApJ}}
\newcommand{\aj}{{ AJ}}
\newcommand{\apjl}{{ ApJL}}
\newcommand{\apjs}{{ApJS}}
\newcommand{\aplc}{{\em Astrophys. Lett. Comm.}}
\newcommand{\prl}{{PRL}}
\newcommand{\mnras}{{MNRAS}}
\newcommand{\araa}{{ ARA\&A }}
\newcommand{\aap}{{ A\&A}}
\newcommand{\aaps}{{ A\&A Supp}}
\newcommand{\pasj}{{ PASJ}}
\newcommand{\nat}{{ Nature}}

\title
[Lensing statistics in surveys]
{Strong lensing statistics in large, $z\lesssim0.2$ surveys: bias in the 
lens galaxy population}
\author
[M\"oller, Kitzbichler, Natarajan]
{
Ole M\"oller$^{1}$, Manfred Kitzbichler$^1$, Priyamvada Natarajan$^{2,3}$\\
$^1$Max-Planck Institute f\"ur Astrophysik, Karl-Schwarzschild-Strasse
1, D-85741 Garching, Germany.\\
$^2$Department of Astronomy, Yale University, P.O.Box 208101, New Haven,
CT 06511-8101, USA\\
$^3$Department of Physics, Yale University, P.O.Box 208120, New Haven,
CT 06520-8120, USA\\
}
\maketitle

\begin{abstract}
Large current and future surveys, like the Two degree Field survey
(2dF), the Sloan Digital Sky Survey (SDSS) and the proposed
Kilo-Degree Survey (KIDS) are likely to provide us with many new
strong gravitational lenses. Taking cosmological parameters as known,
we calculate the expected lensing statistics of the galaxy population
in large, low-redshift surveys. Galaxies are modeled using realistic,
multiple components: a dark matter halo, a bulge component and disc.
We use semi-analytic models of galaxies coupled with dark matter haloes
in the Millennium Run to model the lens galaxy population.
Replicating the selection criteria of the 2dF, we create a mock galaxy
catalogue. We predict that a fraction of $1.4\pm0.18\times10^{-3}$ of
radio sources will be lensed by galaxies within a survey like the 2dF
below $z<0.2$. We find that proper inclusion of the baryonic component
is crucial for calculating lensing statistics -- pure dark matter
haloes produce lensing cross sections several orders of magnitude
lower. With a simulated sample of lensed radio sources, the predicted
lensing galaxy population consists mainly of ellipticals ($\sim80\%$)
with an average lens velocity dispersion of $164 \pm
3\,\mathrm{km\,s^{-1}}$, producing typical image separations of
$\sim3\arcsec$. The lens galaxy population lies on the fundamental
plane but its velocity dispersion distribution is shifted to higher
values compared to all early-type galaxies. We show that magnification
bias affects lens statistics very strongly and increases the 4:2 image
ratio drastically. Taking this effect into account, we predict that
the ratio of 4:2 image systems is $30\pm5\%$, consistent with the
observed ratio found in the Cosmic Lens All-Sky Survey (CLASS). We
also find that the population of 4-image lens galaxies differs
markedly from the population of lens galaxies in 2-image systems. We
find that while most lenses tend to be ellipticals, galaxies that
produce 4 image systems preferentially tend to be lower velocity
dispersion systems with more pronounced disc components.  Our key
result is the explicit demonstration that the population of lens
galaxies differs markedly from the galaxy population as a whole: lens
galaxies have a higher average luminosity and, for a given luminosity,
they reside in more massive haloes than the overall sample of
ellipticals. This bias restricts our ability to
infer galaxy evolution parameters from a sample of lensing galaxies.
\end{abstract}

\section{Introduction}
\label{intro}

Individual lens modeling has already provided many interesting
constraints on the mass profile of clusters and galaxies
\citep[e.g.][]{schechter1997,cohn2001,rusin2001a,wyithe2002,kneib2003,treu2003,broadhurst2005}.
In the case of galaxy lenses, the measured time-delays between the
multiple images has been used to determine bounds on the value of the
cosmological constant \citep[e.g.][]{rhee1991,koopmans2003} and the
Hubble constant \citep{kundic1997a,schechter1997,ferreras2005}.  The
cosmological density parameters, $\Omega_{\Lambda}$ and $\Omega_{m}$
have also been estimated using a sample of gravitational lens systems
\citep{helbig1999,meneghetti2005,dalal2005,chae2003}. However, despite
great efforts, modeling of individual gravitational lens systems has
not yet provided constraints on any of the cosmological parameters
comparable to those from other techniques like the studies of the
Cosmic Microwave Background Radiation. The primary reason for this is
that there are many degeneracies in the lens models which are
still poorly understood. Since the details of the mass models have
considerable effect on the derived cosmological parameters, lens model
degeneracies pose a fundamental limitation in exploiting lensing
\citep{moller1998,wucknitz2002,zhao2003}.

A second possibility to constrain cosmology from lensing is to use
lens statistics. With a large sample of observed lenses, the average
mass properties of the galaxy population become more important and
hence the results are largely independent of individual
deviations in the lens mass distributions. This then also enables 
constraints on the mass properties of the galaxy population as a
whole to be obtained.

Even though cosmology crucially affects lensing statistics
\citep[e.g.][]{turner1984,fukugita1992,kochanek1993,chae2003}, several
effects complicate the connection. The source properties, like source
redshift distribution and luminosity function, together with
magnification bias, all affect lensing statistics strongly.  There is
furthermore, degeneracy between lens models, statistical source
properties, magnification bias and cosmological parameters. Previous
constraints on cosmological parameters from strong lensing are thus
not directly comparable in precision to other techniques like
constraints from experiments such as the \emph{WMAP} space mission
\citep{spergel2006}. It is perhaps more fruitful to take the
cosmological parameters as ``known'' quantities and use lensing
statistics instead to constrain the properties of the lensing galaxy
population. This is the approach that we explore in this work.

The statistics of strong lensing by galaxies has been studied 
by several authors in the past
\citep[]{turner1984,fukugita1992,kochanek1993,moller2001,chae2003}.
Most of these studies calculate a lensing cross section analytically
using a simple lens model along with known mass and luminosity
functions. Others are concerned with the effect of specific details of
the mass models on the lens statistics or properties of the source
population \citep{huterer2005}. On larger scales, the arc-statistics
of galaxy clusters has been used to constrain cosmological parameters
\citep{bartelmann1998,meneghetti2005, dalal2005}. Perhaps due to the limited
sample of galaxy lens systems known, the interest in statistical
lensing on galaxy scales has so far concentrated mainly on predicting the number of strongly imaged
systems using simple lens models. However, the number of known lens
systems is increasing steadily and in the near future many large
surveys are likely to detect a significantly increased number of
strong lens systems. Such a large number of lens systems could
potentially constrain masses of galaxies and provide important
constraints on the mass evolution of galaxies between $z\sim0.1$ and
$z\sim1$. Thus, statistical lensing is potentially a very important
tool for galaxy evolution studies.

Detailed statistical lens studies have so far been hampered by the
lack 
of a clearly defined lens sample. Most known lens systems have
been discovered serendipitously and comprehensive surveys like the Cosmic
Lens All-Sky Survey (CLASS) are rare. Simple estimates of the
lensing probability suggest the fraction of sources at
$z\sim1$ being lensed into multiple images is between $10^{-4}$ and
$10^{-2}$. The largest current survey, the Sloan Digital
Sky Survey (SDSS), contains about 500,000 foreground galaxies and about
100 million background galaxies at higher redshifts, and should thus 
contain at least several thousand lensed galaxies.

In this paper, we use semi-analytic models of galaxy formation
\citep{kauffmann1993,somerville1999,kauffmann2004,bower2006,croton2006} to
predict the statistical properties of a large sample of strong lenses
that would be detected by surveys in the radio, optical and infrared
wavebands. We concentrate here on radio sources that are lensed by low
redshift galaxy lenses with $z_{\mathrm{lens}}<0.2$, as environmental
effects are smaller for such a population. Two large surveys with
similar redshift limits have been carried out to date, the Sloan
Digital Sky Survey \citep[SDSS][]{abdel2006} and the Two Degree Field
Survey \citep[2dF][]{colless2001}. Even though we use the 2dF as a template
survey for this work, our results are equally valid for the SDSS. In
particular, we predict the expected lensing
properties of galaxies in these surveys. This is done by calculating the
lensing properties of a mock survey catalogue obtained from the
semi-analytic models assuming a composite mass model consisting of
dark matter halo (DM), bulge and disc for lens galaxies. The lensing
cross section, image geometries and magnifications are calculated for
each galaxy individually using numerical routines. From these we
then create a sample of lens systems, which we then analyse further.

We describe the use of semi-analytic methods to create the lens mass
models and outline the method to calculate lensing probabilities in
detail in \S\,\ref{method}. In \S\,\ref{maps} we show the predicted
statistical lensing properties of galaxies in low redshift surveys and
the dependence on galaxy properties. In \S\,\ref{classsample} we apply
specific selection criteria and predict the properties of a radio
selected sample of lenses with $z_{\mathrm{l}}<0.2$. Our results are
presented in \S\,\ref{discuss}, including the predictions for the
lensing rate. In \S\,\ref{summary} we summarize our key results and
discuss their implications.  Throughout this paper we use a standard
$\Lambda$CDM cosmology with $\Omega_{\Lambda}=0.75$,
$\Omega_{\mathrm{m}}=0.25$ and
$H_0=73\,\mathrm{km\,s^{-1}\,Mpc^{-1}}$.


\section{The mock lens catalogue}
\label{method}

We use the Millennium Run numerical simulation \citep{springel2005}
together with semi-analytic modelling of galaxy formation to predict
the properties of the dark and luminous matter in galaxies and thereby
derive the expected lensing statistics in large surveys at
$z\lesssim0.2$.  In the procedure used for semi-analytic modeling a
catalogue of halo and substructure is generated at closely spaced
redshift intervals, and for each individual $z=0$ DM halo a
corresponding merger tree is identified. The prescription described in
\citet{croton2006} is used to follow stellar masses, luminosities and
morphologies of galaxies that assemble in these DM haloes. From these,
we create a mock low redshift galaxy catalogue that is similar to the
2dF North field by applying the same selection cuts in magnitude and
redshift. We describe the creation of this mock 2dF galaxy catalogue
below. The parent DM haloes are taken from the Millennium Run. Selection
of simulated haloes is done by defining a backward lightcone from $z =
0$. Even though we use the creation of a backward lightcone, we note
that for most of the work presented here, the creation of a full light
cone is not strictly necessary, as we treat galaxies as isolated in
our lensing analysis. In this treatment, we are ignoring the immediate
environments of haloes while calculating lensing cross sections and
image geometries. It is well documented that the presence of mass
concentrations in the vicinity of lenses impacts the image separation
distribution\citep{moller2001} and we address this interplay in future
work.

\subsection{Making mock observations}
\label{subsec:lightcones}

We construct mock observations of our artificial Universe in the
Millennium simulation by positioning a virtual observer in the
simulation at zero redshift and finding those galaxies which lie on
the observer's backward lightcone. The main limitation in producing a
mock observation of a simulation is the finite box size, which in this
case is $500\,h^{-1}\rm{Mpc}$ on a side.  The periodic nature of the
simulation allows us to fill the space with any number of boxes we
require but we still have to deal with periodic replications that
would appear if we were to look through the simulation volume along
one of its preferred axes. We can avoid this kaleidoscopic effect
through slanting the observed cone by a certain angle.  After having
determined the survey geometry, observer position and line-of-sight we
fill the four-dimensional Euclidean space-time with a grid of
simulation boxes. In the three spatial coordinates we make use of the
periodicity of the simulation whereas the time coordinate is given by
the 64 snapshot times corresponding to their respective output
redshifts. In practice, only those cells in the space-time grid are
populated with galaxies which actually intersect the backward
lightcone in the observed field-of-view.  After coarsely filling the
volume around the observed lightcone with simulation boxes in this way
one can simply ``chisel off'' the protruding volume.

We note that by using comoving coordinates and assuming a flat
Universe we have the luxury of cutting the lightcone from our
simulated volume simply like we would in Euclidean geometry. In
general this is a non-trivial endeavour that requires taking into
account the curvature of the Universe as well as its expansion with
time.

\subsection{Lensing probabilities for an SIS model}

Before calculating the lensing cross section for realistic mass models
of lensing galaxies we first discuss the singular isothermal sphere
(SIS) model that has in the past most commonly been used for
statistical lensing calculations
\citep[e.g.][]{turner1984,fukugita1992}.  Despite its simplicity the
SIS model has been surprisingly successful in describing the overall
statistical properties of galaxy-scale lenses. As shown 
by \citet{koopmans2002a,koopmans2003}, the stellar
component of lens galaxies has a density profile that is much steeper
at than the dark matter at small radii so that the total mass profile (including dark matter)
follows an isothermal profile on scales probed by galaxy strong
lensing more closely, thereby explaining the modelling success of the
SIS.  For most groups and cluster lenses, however, the SIS model is an
inadequate description of the total mass distribution. On cluster
scales the baryonic component becomes less important and the lensing
profile is more accurately modeled using a Navarro,
Frenk \& White profile \citep[NFW][]{navarro1997} well beyond
the arc radius \citep{kneib2003}. For lower redshift lenses, the
Einstein radius is small compared to the galaxy size and the relative
DM mass content within the Einstein radius is smaller. Thus, for all
galaxies (excluding the most massive groups and clusters), the
baryonic galaxy component of the mass profile dominates the lensing
cross section at low redshifts.

The lensing cross sections of a SIS model for elliptical galaxies can
be calculated directly from the observed Faber-Jackson relation. The
lensing cross section of an SIS is given by:
\begin{equation}
\tau_{\rm SIS}= 64\pi\left(\frac{\sigma}{c}\right)^4\left(\frac{D_{\mathrm{LS}}}{D_{\mathrm{OS}}}\right)^2, 
\end{equation}
the total lensing probability can be calculated using the luminosity
function $N(L)$ of galaxies:
\begin{equation}
\tau\propto\int_0^{\infty}\sigma(L)^4N(L)\,dL,
\end{equation}
where 
\begin{equation}
\sigma(L)=\sigma_*\left(\frac{L}{L_*}\right)^{\beta}
\label{lf.eq}
\end{equation}
is the Faber-Jackson relation. Using the 2dF 
luminosity function \citet{norberg2002} fit, we have:
\begin{equation}
\tau=\int_{z=0}^{\infty}\,\tau_*\,\left(\frac{D_{\mathrm{LS}}}{D_{\mathrm{OS}}}\right)^2\,\phi_*\left[\frac{\sigma}{\sigma_*}\right]^4\,\left(\frac{L}{L_*}\right)^{\alpha}\,\exp{(-\frac{L}{L_*})}\frac{dV_{\mathrm{co}}}{dz}\,dz
\end{equation}
where $\phi_*=1.64\times10^{-2}\,\mathrm{Mpc}^{-3}$, $\alpha=-1.21$,
$L_*=1.82\times10^{10}L_{\odot}$ and
$\tau_*=(\sigma_*/\mathrm{km\,s^{-1}})^4\times1.055\times10^{-9}\,\mathrm{arcsec}^2$
is the characteristic lensing cross section of an $L_*$ galaxy for
equal distances $D_{\mathrm{LS}}$ and $D_{\mathrm{OS}}$. The comoving volume
of a shell of thickness $dz$ is given by $dV_{\mathrm{co}}/dz\times
dz$.

\begin{figure}
\epsfig{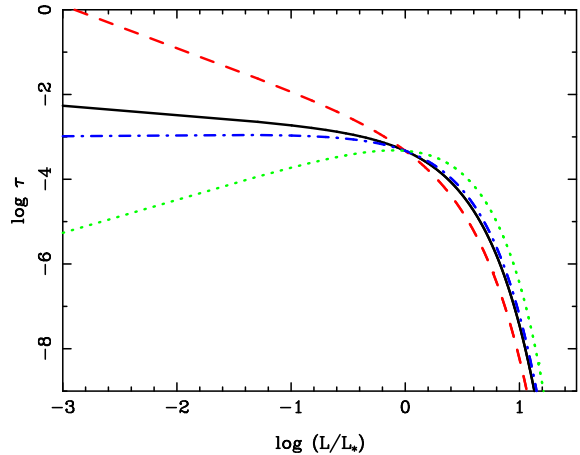}
\caption{The lensing cross section for SIS lenses with sources at
  $z = 2$ for different slopes of the luminosity function and
  power-law indices relating luminosity and velocity dispersion. 
  The solid line shows the results for $\alpha=-1.21$
  and $\beta=0.25$, corresponding to the 2dF results. The dashed, red
  line is for $\alpha=-2$ and $\beta=0.25$ and the dotted green and
  dot-dashed blue lines are  for $\alpha=-1.21$ and $\beta=0.31$ and
  $\beta=0.5$ respectively. The functional form of the luminosity
  function is given in equation \ref{lf.eq}.}
\label{sisfig}
\end{figure}

We show the $\tau-L$ relation for various choices of $\alpha$ and
$\beta$ in Fig.\,\ref{sisfig}. For a slope of the Faber-Jackson relation of
$\beta < 0.3$ the lensing cross section $\tau(L)$ is a monotonically
decreasing function with L. In the case of an SIS lens, the largest
contribution to the total lensing cross section is expected from the
galaxy population around $L_*$.  Using the Faber-Jackson relation
determined from early-type galaxies in the SDSS \citep{bernardi2003c},
with $\beta=1/4$ and $\sigma_*=\sigma(L_*)=166.44\,\mathrm{km\,s^{-1}}$,
we obtain a total lensing cross section of:
\begin{equation}
\tau_{\mathrm{tot}}=\int_{L=0}^{\infty}\tau(L)\,dL=8.36\times10^{-3},
\end{equation}
for sources at $z_{\rm sources}=2$. This is our first and simplest
estimate of the expected lensing rate. At this juncture, we have
neglected the fact that galaxies contain DM and baryons which are, in
general, not expected to be distributed isothermally. The measured
velocity dispersion may be biased with respect to the velocity
dispersion inferred for the dark matter. In addition, we have also
neglected magnification bias and the source luminosity and redshift
distributions.

\subsection{Towards realistic lens models: dark matter and baryons}

More sophisticated models than the SIS have been used by e.g.
\citet{oguri2005} who predicted the lensing statistics in the SDSS
using a spherical Hernquist model for the baryons and an NFW dark
matter mass profile. Spherical models do predict the overall lensing
cross section with adequate accuracy \citep{fukugita1992}. They
cannot, however, predict any detailed properties, like the average
magnification and the statistics of image geometries. For instance,
the number of four image systems predicted by spherical lens models is
zero. We use the ray-tracing code \emph{gLens} \citep{moller1998}
\footnote{The gLens software was developed by M{\"o}ller to solve
the lens equation on an adaptive grid on the source plane. This
ray-tracing code permits the addition of an arbitrary number of mass
components for lenses.}  to calculate the lensing properties of
multi-component lens systems efficiently.  In order to calculate
expected lensing properties of galaxies in large surveys, we need to
relate the observed properties, like luminosity and size to a mass
model. Since lensing is sensitive to the total mass, we need to
include the contributions both of the baryonic and of the dark matter.
Our galaxy model therefore consists of a baryonic component, modeled
as a bulge of mass $M_{\mathrm{b}}$ plus a disc of mass
$M_{\mathrm{d}}$ that reside in a DM halo of virial mass
$M_{\mathrm{vir}}$.

In this paper, we consider only the lensing statistics of isolated
lenses. In reality, most lens galaxies are expected to reside in dense
environments, in groups or clusters
\citep{kundic1997b,moller2002,keeton2004,oguri2005}. However, the lens
galaxies we consider here are all at relatively low redshifts, where
the effect of environment is expected to be less strong. At redshifts
$z \leq 0.2$ the expected Einstein radius is small compared with the
distance to nearest neighbours (this is true even for lens galaxies in
groups and clusters). We note that by construction our catalogue also
includes the total lensing cross section by groups and clusters: the
haloes of these objects, together with their central galaxy, are
included in our catalogue as single, isolated galaxies. In this sense,
our calculations are complete, except that they do not take into
account the effects of satellite galaxies and external shear deriving
from neigbouring mass concentrations and halo substructure within
these DM haloes.

\subsubsection{Modeling the dark matter halo}

N-body simulations of structure formation in a cold dark matter
dominated Universe find that the best-fit density profile for dark
matter haloes follows a universal functional form over a broad range
of scales \citep{navarro1997}.  The mass distribution for a spherical
NFW halo is given by,
\begin{equation}
\rho(r)=\frac{\delta_c\rho_c(z)}{\frac{r}{r_s}\left(1+\frac{r}{r_s}\right)^2},
\end{equation}
where $r_s=R_{\mathrm{vir}}/c$ is the scale length, $\rho_c(z)$ is the critical
density for closure at the galaxy redshift and
\begin{equation}
\delta_c=\frac{200/3\,c^3}{\log{(1+c)}-c/(1+c)}.
\end{equation}
The concentration parameter $c$ is ultimately a unique property for
each halo in simulations, however correlations (with some scatter)
between the concentration parameter and the mass or virial velocity of
the halo have been found \citep{navarro1997,bullock2001,eke2001}. A
closer investigation of the strong lensing by NFW haloes reveals that
the lensing cross section depends crucially on the concentration
parameter. For the same halo mass, a halo with $c=20$ has a lensing
cross section that is $\sim$10 times higher than a halo with $c=10$,
and $\sim100$ times higher than a halo with $c=5$. This implies that
the scatter in these correlations between halo mass and central
concentration parameter is very important for lensing -- most of the
lensing cross section contribution derives from massive, high
concentration `outliers'. Due to this effect, the lensing statistics
of a galaxy population can only be correctly predicted if the full
scatter in the distribution function of the concentration parameter is taken into account. In this
work, we calculate the concentration parameter explicitly for each
halo. For this, we use the equation \citep{navarro1997},
\begin{equation}
\frac{V_{\mathrm{max}}}{V_{\mathrm{vir}}}=\sqrt{0.216\frac{c}{A_{\mathrm{c}}}},
\end{equation}
where
\begin{equation}
A_{\mathrm{c}}=\ln{(1+c)}-\frac{c}{1+c}.
\end{equation}
For each halo in the simulation, the parameters $V_{\mathrm{max}}$ and
$V_{\mathrm{vir}}$ are known and we use the eqns. (8) and (9) to
calculate the concentration parameter for each halo. This procedure
ensures that the distribution of concentration parameters
automatically has the correct scatter.

\subsubsection{Modeling the baryonic components}

The baryonic mass is distributed into a bulge and a disc component.
The relation between the mass of a DM halo and the luminosity of the
baryonic component can be studied using semi-analytic models of galaxy
formation. The semi-analytic prescription of \citet{croton2006} gives
a bulge luminosity, a disc luminosity and a disc size.

\begin{figure}
\epsfig{file=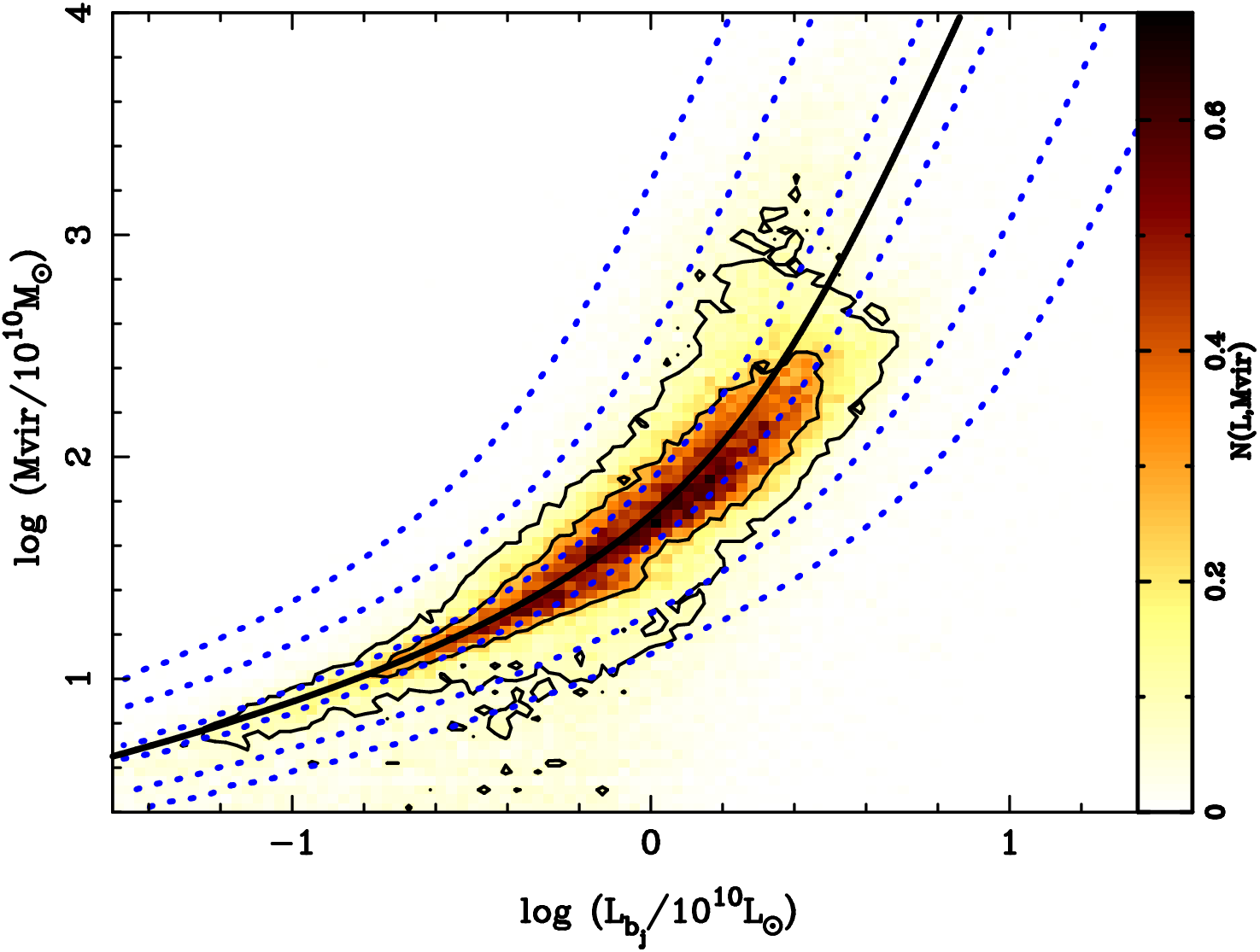, width=6.0cm, angle=-90}
\caption{The relation between the halo virial mass $M_{\mathrm{vir}}$  and
  galaxy luminosity in the $B_j$-band. The
  shading shows the number of galaxies in the mock 2dF catalogue 
   created from the
  Millennium Run \citep{springel2005} combined with the semi-analytic model of
  \citet{croton2006}. The black line shows the
  relation in the $B_j$-band obtained by \citet{vale2004}. The dotted blue lines
  show the 68, 90 and 99 percent levels of the scatter around the
  mass-luminosity relation as obtained by \citet{cooray2005}.}
\label{valefig}
\end{figure}

We show the mass-luminosity relation for our mock catalogue in
Fig.\,\ref{valefig}.  For comparison, we overplot the relation
\begin{equation}
L_{Bj}/10^{10} L_{\odot}=0.3\frac{(M_{\mathrm{vir}}/10^{11}M_{\odot})^4}{\left(0.57+(M_{\mathrm{vir}}/10^{11}M_{\odot})^{0.79}\right)^{1/0.23}}.
\end{equation}
derived by \citet{vale2004} as the solid line, including the scatter (dotted,
blue line). Note that there are galaxies in the catalogue that
consistently fall above and to the left of the analytic relation
(given in eqn. 10) for high virial masses, which correspond to central
galaxies in the more massive haloes of clusters or groups.

\subsubsection{The Bulge component}
\label{bulge}

The size and velocity dispersion of the bulge component is not
determined in the semi-analytic modelling. However, the distribution
of bulge sizes is known observationally \citep{dejong2004}. There are
no models that relate the bulge size to properties of the DM halo.
The fundamental plane relation suggests that DM haloes and the bulge
properties are related: virialised haloes with constant M/L will
naturally lie on the fundamental plane, but so far no theoretical
model has been developed that naturally explains both the existence
and the slope of the fundamental plane. In the absence of such a
model, we rely on the observed relations between luminosities and
sizes of elliptical galaxies and apply these to the population of
bulges as a whole, irrespective of the morphological type of the
galaxy.  In the SDSS, \citet{bernardi2003c} find a correlation between
half-light radius, $R_{\mathrm{b}}$ and galaxy luminosity of the form:
\begin{equation}
\log{R_{\mathrm{b}}}-0.52=\frac{1}{\alpha}\log{(L/L_*)},
\end{equation}
where $\alpha=1.5\pm0.06$ and $L_*=1.58\times10^{10}L_{\odot}$. They also
find the Faber-Jackson relation for the velocity dispersion
\begin{equation}
\log{\sigma}-2.197=\beta\log{L/L_*},
\end{equation}
where $\beta=0.25\pm0.06$. Since the semi-analytic model fixes the total
bulge luminosity and DM properties, we do not use the latter
relation. Instead, we only use the $R_{\mathrm{b}}-L$ relation
(eqn. 11 above) to determine the bulge-size and then solve the
Jeans-equation to obtain the line of sight velocity dispersion
$\sigma$, assuming a constant mass-to-light ratio of $\Gamma=7$ in the
B-band, which
is the mean M/L ratio expected for bulges and early-type galaxies \citep{gerhard2001,cappellari2006}. This choice
does not affect the tightness of the fundamental plane relation
strongly; the uncertainty in $\alpha$ is the strongest contributor to
the scatter in the relation. We can then check whether the bulges
still satisfy the observed $\sigma-L$ relation and fall on the
fundamental plane consistent with observed bulge properties.
  
\begin{figure}
\epsfig{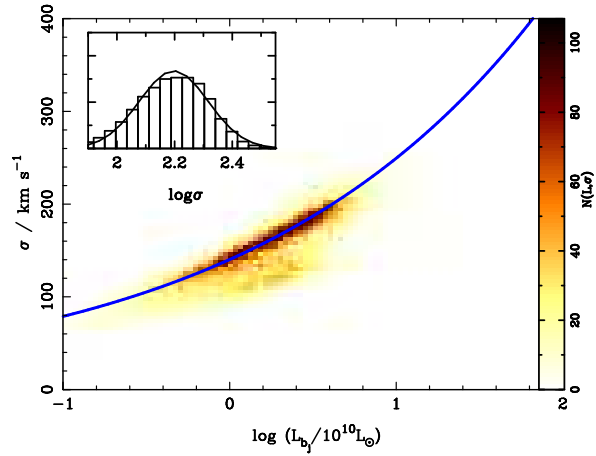}
\caption{The Faber-Jackson relation: the number
  of galaxies in the mock catalogue as a function of $L$ and
  $\sigma$ are shown, with the blue
  line showing the best-fit observed relation $\sigma=\sigma_*(L/L_*)^{0.25}$.
  The inset shows the distribution of velocity dispersions,
  $\log\sigma_{\mathrm{b}}$ in our mock 2df catalogue. The smooth line
  in the inset shows the observed velocity dispersion distribution in
  the SDSS found by \citet{bernardi2003a}.
}
\label{vdistfig}
\end{figure}
\begin{figure}
\epsfig{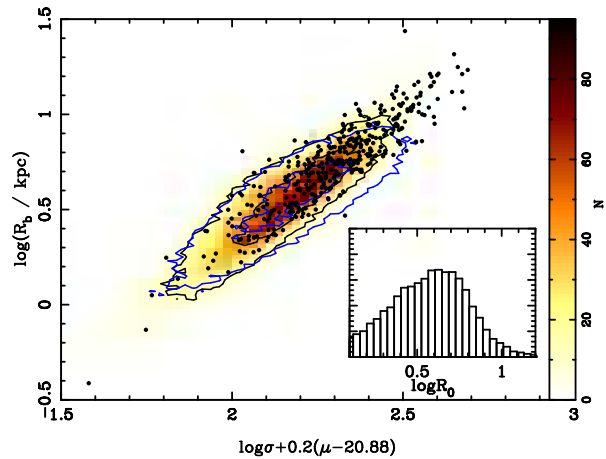}
\caption{The fundamental plane for galaxies in the mock catalogue: the number 
of galaxies in the mock as a function of
  $R_{\mathrm{b}}$ and a combination of velocity dispersion and
  surface brightness.  The blue contours show the same for the sample of
  elliptical galaxies in the SDSS \citep{bernardi2003a}. The histogram
  in the inset shows the distribution of $\log{R_{\mathrm{b}}}$ in the
  mock catalogue. The dots show the mock lens sample created in \S\,\ref{classsample}.
}
\label{rdistfig}
\end{figure}

We show the Faber-Jackson relation and the histograms of the
distributions of the velocity dispersion and bulge size in
Figs.\,\ref{vdistfig}\,\&\,\ref{rdistfig} for the mock catalogue.
These plots show that, using only the observed $R_{\mathrm{b}}-L$
relation as input, we recover the observed Faber-Jackson relation and
the fundamental plane. Of particular importance is the fact that we
recover the observed velocity dispersion distribution measured for
early-type galaxies by the SDSS. We conclude therefore that
our mock galaxies are realistic.

These structural parameters are then used to model the bulge component
in the lens as a de Vaucouleur profile \citep{devaucouleurs1948},
\begin{equation}
\Sigma_{\mathrm{b}}=\Sigma_{0,\mathrm{b}}\times \exp{-(r/r_{\mathrm{s}})^{1/4}},\end{equation}
where $r_s=3681\times R_{\mathrm{b}}\times 0.551$ and
$\Sigma_{0,\mathrm{b}}=\Gamma L/8\pi\times
5040r_{\mathrm{s}}^2\sqrt{1-e^2}$, measured in $\mathrm{kpc}$ and
$\mathrm{M_{\odot}kpc^{-2}}$ respectively, and where $\Gamma$ is the 
fixed mass-to-light ratio of bulges taken to be $7 M_{\odot}/L_{\odot}$ in the B-band.
We also allow for ellipticity, $e$, in the bulge. To assign ellipticities to the
bulge component in our mock galaxy catalogue, we draw an
ellipticity randomly from a Gaussian distribution of mean
$\average{e}=0.3$ and standard deviation, $\sigma_e=0.3$, consistent with
the observed ellipticities of early-type galaxies \citep{jorgensen1994}.

\subsubsection{Disc component}

In the semi-analytic models of galaxy formation every galaxy,
irrespective of morphology, also contains a disc component. The infall
and merger history of haloes determines the disc properties: total disc
mass, $M_{\mathrm{d}}=\Gamma L_{\mathrm{d}}$ and disc size,
$R_{\mathrm{d}}$. The mass distribution of discs is taken to be
well-fit by an exponential:
\begin{equation}
\Sigma_{\mathrm{d}}=\Sigma_0\exp{(-r/R_{\mathrm{d}})},
\end{equation}
where $\Sigma_0=M_{\mathrm{d}}/\pi R_{\mathrm{d}}^2$ is the central
surface mass density of the disc.
\begin{figure}
\epsfig{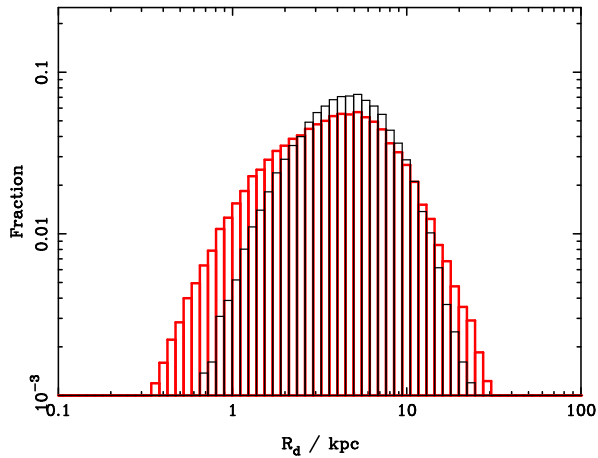}
\caption{The distribution of disc sizes in the mock 2dF
  catalogue. The histograms show the distribution of
  $\log R_{\mathrm{d}}$ in the mock, the thick, red is for all galaxies
and thin, black is for spiral galaxies only. Note that both histograms have
 been normalised independently.} 
\label{rddistfig}
\end{figure}

We plot the distributions of the structural disc parameters for the
mock catalogue in Fig.\,\ref{rddistfig}. Identifying the maximum
velocity dispersion $V_{\mathrm{max}}$ of the DM halo with the disc
rotational velocity, $v_{\mathrm{c}}$ also gives the Tully-Fisher
relation for galaxies in our mock catalogue (see Fig.~6).

\begin{figure}
\epsfig{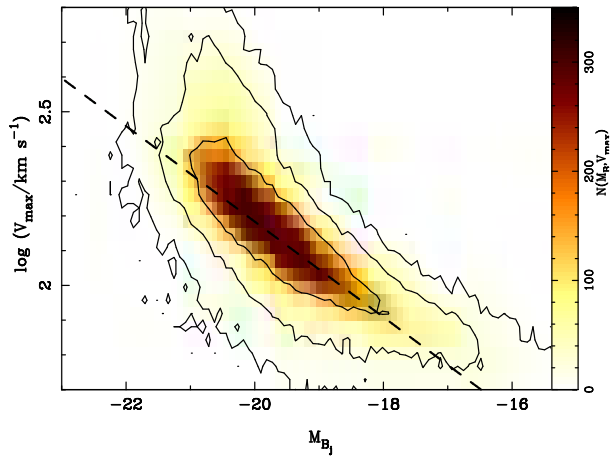}
\caption{The distribution of number of galaxies as a function of
  $V_{\mathrm{max}}$ and disc luminosity in the mock
  catalogue. The dashed line shows the observed Tully-Fisher relation,
  $M_{\mathrm{B_\mathrm{j}}}=-4.125-7.27\times \log{V_{\mathrm{max}}}$\citep{tully1977,pierce1992,tully2000}.}
\label{tully.fig}
\end{figure}

The blue line in Fig.\,\ref{tully.fig} shows the Tully-Fisher relation
observed by \citet{tully2000} and \citet{pierce1992}. There is a very
good match between observed and predicted relation for
$M_{\mathrm{B_j}}\gtrsim -21$, corresponding to
$L\lesssim2.15\times10^{10}L_{\odot}$. At higher luminosities, there
is an upturn in the predicted relation. Given the small observed
number of objects with measured rotational velocities at these
luminosities this is consistent with observational data at the present
time.

\subsection{Lensing properties of the composite model}

To illustrate the general lensing properties of the composite mass
model used, we show the image and source magnification maps of the
dark matter+bulge+disc models in
%
%
\begin{figure*}
\epsfig{file=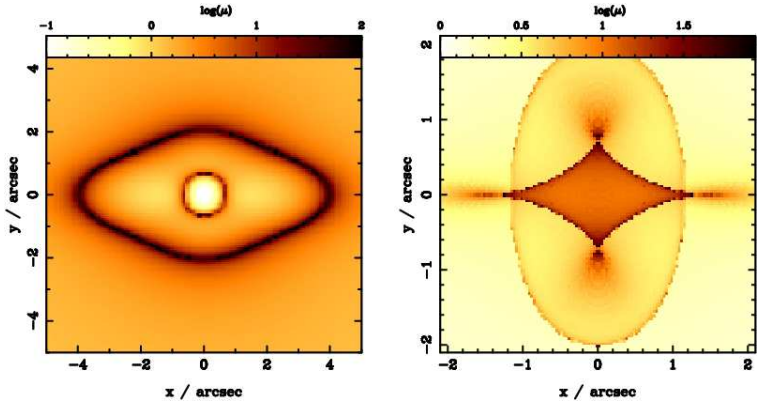, width=10.0cm, angle=0}
\caption{The image (left) and source (right) magnification maps for dark matter+bulge+disc model of a
  luminous $L=1.8\times10^{11}L_{\odot}$ galaxy. The colour scale shows
  the magnification of a point source as a function of image and source position
  respectively. The source plane is at
  $z_{\mathrm{s}}=1$ and the image plane is at 
  $z=0.15$. The lens has a bulge to total mass ratio
  B/T of $0.2$, a bulge ellipticity of $e=0.4$ and a disc
  inclination angle of $\theta=70\,\deg$. The dark halo mass is
  $M_{\mathrm{vir}}=1.3\times10^{13}M_{\odot}$ inside
  $R_{\mathrm{vir}}=455\,\mathrm{kpc}$, with a concentration parameter
  of $c=9$.}
\label{lensingprops.fig}
\end{figure*}
%
%
Fig.\,\ref{lensingprops.fig} for a spiral galaxy with
$L=10^{11} L_{\odot}$. Note the strong asymmetry introduced by the
disc component. This is expected since for inclination angles of
$\theta>65\,\deg$, the projected surface mass density along the major
axis is strongly increased for thin discs \citep{moller1998}.

\begin{figure}
\epsfig{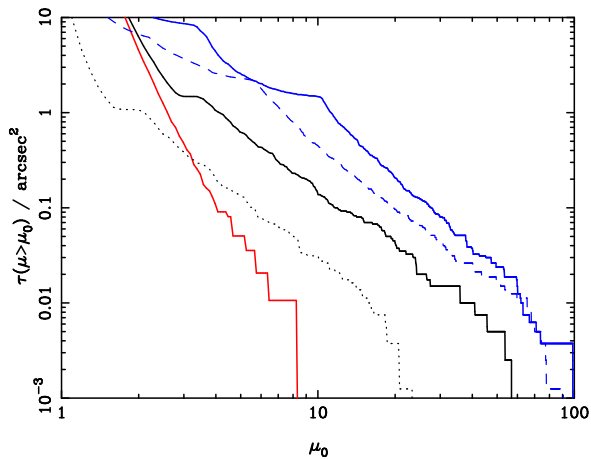}
\caption{The lensing cross section as a function of magnification
  $\mu$ for DM, DM+bulge and DM+bulge+disc models of a
  typical $L=10^{11}L_{\odot}$ galaxy in the mock catalogue. Cumulative
  histograms of the cross sectional area for lensing of point sources
  by magnifications of more than $\mu_0$, $\tau(\mu>\mu_0)$ are shown. The solid, red
  line is for the pure DM model. The solid, black line shows the result 
of adding a 
  bulge and the solid blue line is for the complete DM+bulge+disc
  model. The dashed black and blue lines show the magnification cross sections for
  model of bulge and bulge+disc without dark matter, respectively. Model parameters are as for the previous figure.}
\label{lensingcross.fig}
\end{figure}

The magnification curves in the source and image planes are shown in
Fig.\,\ref{lensingcross.fig}. The NFW dark matter profile on its own does not
have a significant lensing cross section in this example, as the
concentration parameter $c$ of the dark matter halo is too low. It is
interesting to note, however, that despite this the magnification
cross sections for the composite model are influenced
significantly by the DM component. This can be seen by comparing the
lensing cross-sections as a function of magnification shown as solid
(DM) and dashed (no DM) lines in Fig.\,\ref{lensingcross.fig}. The
cross section for the case with no DM is a factor of $\sim2$ below
that with DM, for almost all magnifications above about
3. Therefore, even in cases when the DM halo is not concentrated
enough to produce lensing, it still affects the lensing cross section
significantly.

\section{The expected lens properties of the galaxy population in the 2dF}
\label{maps}
\subsection{Calculation of lensing properties}

Once each galaxy has been assigned a NFW halo+bulge+disc mass model,
the calculation of the lensing probability for a given source
population depends only on the geometric parameters, i.e. the redshifts
of the source, lens and the cosmological parameters. We are not
interested in constraining cosmological parameters with lensing in
this work and we fix the underlying cosmology as described in
\S\,\ref{intro}.

We calculate the lensing properties of each of the $\sim70,000$ galaxies in
the mock catalogue using ray-tracing. Since we consider only one lens
plane, this reduces to numerically solving the lens equation,
\begin{equation}
\bf{\theta}=\bf{\beta}-\bf{\alpha},
\end{equation}
where $\bf{\theta}$ is the imaged position of a source at angular position
$\bf{\beta}$ and $\bf{\alpha}$ is the deflection angle.
 
For the calculations in this section, we fix the sources at
$z_{\mathrm{source}}=1$. In order to allow ray-tracing calculations of
sufficient accuracy in a reasonable amount of time we use the adaptive
grid technique, as described in \citet{moller2001}. For each galaxy,
an approximate lensing region size is determined using the extent of
the Einstein radius. This is done by finding the points in the image
plane at which the magnification diverges. The physical size of the
putative lensing region is then set to $20$ times the distance of the
most distant such points to the galaxy, giving a conservative upper
estimate of the region within which multiple imaging can occur.  This
region in the image plane is then covered with an initial grid of
$40\times40$ rectangles. Each grid point is then mapped to the source
plane. After mapping, magnifications on the image plane are given by
the ratio of the areas of the original to that of the mapped
rectangles. Depending on the magnification $\mu$ the initial grid is
then adaptively refined by subdividing each rectangle into
$n_{\mathrm{sub}}$ smaller rectangles, with
$n_{\mathrm{sub}}(\mu)=1+(\mu-5)/195\times20$, for $\mu>5$, and with
no refinement otherwise. This resulting adaptive grid is then also
mapped onto the source plane. Finally, the image magnifications and
image multiplicities are calculated on the resulting fine grid on the
source plane.

\subsection{Total lensing cross sections and image separation
  distributions}

Two statistics have been used in the past to constrain cosmological
parameters \citep{chae2003,kochanek1993} and lens galaxy population
properties \citep{chae2005,kochanek1996}: total lensing cross sections
and image separation distributions. The total lensing cross
section we obtain from our mock catalogue is
$100752\,\mathrm{arcsec}^2$ in a total survey area of $657\deg^2$,
with 79042 galaxies below redshift of 0.2 constituting our lens galaxy
population. This gives a total probability of a source being lensed as
$1.13\times10^{-5}$. This estimate neglects magnification bias and we
show in \S\,\ref{classsample} that magnification bias increases the
lensing probability dramatically.
 
We plot the lensing cross section for our mock catalogue as a function
of lens galaxy luminosity and halo virial mass in
\begin{figure}
\epsfig{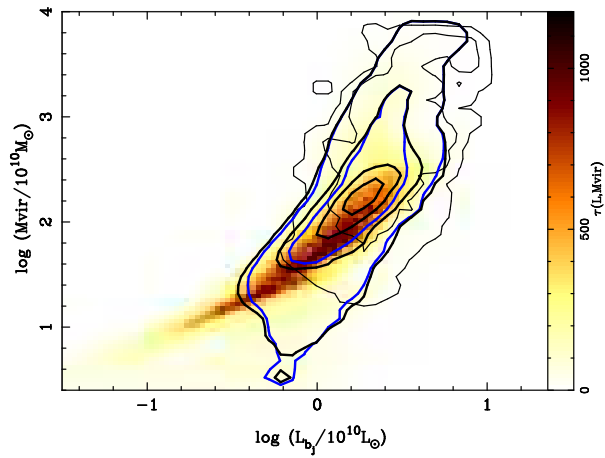}
\caption{The lensing cross section as function of luminosity and
  mass. The colour scale shows the distribution of all galaxies as a
  function of $L$ and $M_{\mathrm{vir}}$. The blue contour shows the
  lensing cross section distribution for the complete DM+bulge+disc
  models. The thick, black contour shows the cross section for
  DM+bulge models and the thin black contours are for DM only
  models. The contour levels are $100$, $300$, $700$ and
  $1000\,\mathrm{arcsec}^2$ for the first two cases and $0.1$, $1$ and $10\,\mathrm{arcsec}^2$ in
  the latter case.}
\label{lumcross.fig}
\end{figure}
\begin{figure}
\epsfig{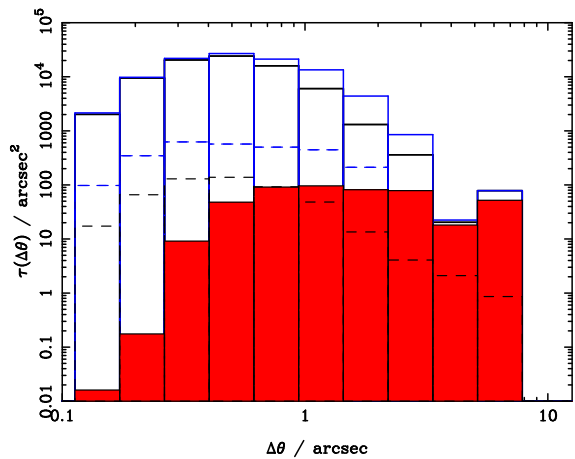}
\caption{The expected image separations in the 2dF. As in the previous
  figure, we show the results for the DM only (red), the DM+bulge but no disc (black) and the
  DM+bulge+disc (blue) models. The solid lines correspond to the image
  separation histogram of 2 image systems, the dotted line shows the
  maximum image separation for 4 image systems. In all cases, shown above the
  effect of magnification bias is not included.}
\label{erdist.fig}
\end{figure}
%
Fig.\,\ref{lumcross.fig}. The contours show the lensing cross section
for background sources at $z_{\mathrm{source}}=1$ as a function of
mass, $M_{\mathrm{vir}}$ and luminosity $L_{\mathrm{B_j}}$. Comparison
with the underlying distribution for all galaxies in the sample shows
that the contribution to strong lensing cross section comes
preferentially from massive, luminous galaxies. The shift in the
luminosity from a peak luminosity of
$L^{\mathrm{peak}}_{\mathrm{tot}}\sim1 0^{10}L_{\odot}$ for the total
distribution to a peak of $L^{\mathrm{peak}}_{\mathrm{lens}}=2\times
10^{10}L_{\odot}$ for the lensing sample is clearly strong. An
additional shift to higher virial masses by a factor of $\sim 2$ 
is also evident. The baryonic component dominates the lensing cross
section of galaxies, but, for a fixed luminosity, the contribution of
high mass objects to the lensing cross section is larger. The dark
matter component therefore also plays a key role.  In
Fig.\,\ref{erdist.fig} we show the expected distribution of image
separations. Here, we clearly see the relative effects of the various
lens components. For small image separations (corresponding to low
masses) the cross section is completely dominated by the baryonic
component. Large image separation systems, mainly corresponding to
groups and clusters, have lensing cross sections that depend much more
on the dark matter content.
 
It is clear from these plots that the baryonic component is absolutely
crucial in determining statistical lensing properties on galaxy
scales.  The importance of the baryonic contribution to the lensing
cross sections of DM haloes is clearly seen: it increases by a factor of $10-1000$
for systems with image separations between $\sim0.1$ and
$\sim1\,\arcsec$. The increase is smaller for systems with large image
separations. These systems correspond to lensing by massive central
cluster and group galaxies. As the Einstein radius increases for more
massive systems, the fraction of DM within the Einstein radius also
increases, due to the shallower slope of the DM component. Despite
this importance of the baryonic component, the lensing cross section at a fixed $L$
does depend on the DM halo mass, $M_{\mathrm{vir}}$, even for highly
luminous galaxies (cf. Fig.\,\ref{lensingcross.fig}).

\subsection{Image geometries: two and four image systems}

Previous work by e.g. \citet{moller1998} and \citet{blain1998} has
shown that baryonic components can significantly change the total
lensing cross section. The disc component plays a major role in this;
inclined discs can increase the lensing cross section
significantly. However, as suggested by \citet{moller2003} this effect
is important only for late-type systems, whereas early-type galaxies
have too small a disc component to change the lensing cross section
significantly. The ratio of 4 to 2 image systems
\footnote{Note that strong lensing produces an odd number of images,
one of the images is always strongly de-magnified, therefore 2 images
systems have a total of 3 images with 2 magnified ones and one central
demagnified one and likewise a 4 image system has in fact a total of 5
images.}, however, may be affected strongly, as even relatively low mass discs
can increase the asymmetry of the lensing potential noticeably.

\begin{figure}
\epsfig{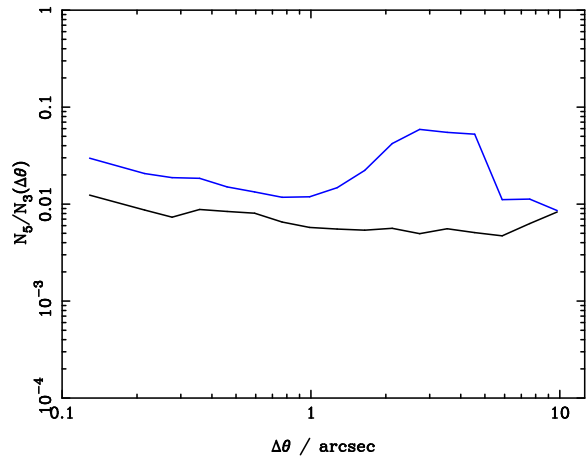}
\caption{The average ratio of the lensing cross section for 4: 2
  image-systems as a function of image separation. The black
  line is for a DM+bulge only model for all galaxies in the mock
  catalogue, and the blue line for a DM+bulge+disc
  model. Magnification bias is not included in this calculation.}
\label{imageratio.fig}
\end{figure}

We show the ratio of 4 to 2 image systems in the catalogue for the
DM+bulge and the DM+bulge+disc models in Fig.\,\ref{imageratio.fig}.
For systems with $10^{11}L_{\odot}\lesssim L\lesssim10^{12}L_{\odot}$
the fraction of late-type galaxies is high enough to increase the
cross section for 4-image systems by more than a factor of
$\sim10$. Again, this result neglects the effect of magnification
bias. Magnification bias is expected to favour late-type galaxies as
4-image lens systems, given that the expected cross section for
magnification of $\mu\sim10-20$ is a factor of up to $\sim10$ larger
for inclined spiral lenses \citep{moller1998}. However, as we show in
\S\,\ref{classsample}, this is not the case: elliptical galaxies have
a smaller cross-sectional area for magnifications in that range, but
they generally have a larger cross section for magnifications above
$20$. This implies that magnification bias in fact favours the
moderately elliptical gravitational potential of early-type galaxies.

\subsection{Selection biases}

The results presented above do not include the effects of selection
biases. All surveys are by construction limited in flux, resolution or
some other observational criteria. These varied selections can
strongly affect the inferred properties of lens samples. A flux
limited survey for example, will be strongly affected by magnification
bias for sources with a steep luminosity function. Since galaxies and
radio sources follow quite steep luminosity functions magnification
bias is important for lensing in the radio as well as in the
optical. One can estimate the effect of magnification bias for simple
selection criteria. The cumulative probability distribution function
for magnification above a value $A$ is generally $P(\mu>A)\propto
A^{-\alpha}$, where $\alpha=2$ for SIS lenses. Assuming that the
probability that a source enters the survey is 1 for $\mu\times F>F_0$
and 0 otherwise, it follows that the fraction of sources of luminosity
$F$ that are observed at flux $F'$ is
$N(F')=\int_0^{\infty}dP/dA(A)N(F'/A)dA$ [cf. \citet{maoz1993} and
\citet{king1996}].  For a lens population with a more complicated lens
model, an analytic calculation of the effect of magnification bias
becomes infeasible. Including the selection effects due to finite
resolution limits requires a numerical calculation -- which we
describe in the next section.

\section{A mock sample of lens systems}
\label{classsample}

We use the lensing code \emph{gLens} together with our mock galaxy
catalogue to create a sample of lens systems, including observational
selection effects. We include two key selection effects: a flux limit
and a resolution limit. Concentrating on lensed radio sources, we
create a sample that has selection criteria in concordance with the
CLASS survey \citep{chae2003}:
\begin{enumerate}
\item The total flux F of each system of images is above $F_0$,
\item the image separation is $>0.03\arcsec$,
\item and for double image systems the magnification ratio of bright
  to faint image is $>0.1$.
\end{enumerate}
We create the sample using rough estimates of the lensing cross
section for each galaxy assuming spherical symmetry, by solving the
lens equation for the approximate Einstein radius,
$\theta_{\mathrm{er}}$. We then populate the total estimated cross
section area, $A_{\mathrm{attempt}}$ with $N_{\mathrm{attempt}}$
sources. For each source, we pick a random flux from a distribution
$N\propto F^{-2.1}$ and assign a source redshift from a redshift
distribution that is a Gaussian with mean
$\average{z_{\mathrm{s}}}=1.27$ and width $0.95$ as found by
\citet{willott2001} in the 6CE and 7CRS radio samples.  
For each source, we find
the positions of the images using an adaptive lensing technique
similar to the one described in \citet{wucknitz2004}. We then apply
the selection criteria (i)-(iii) to the lens system. If the system
satisfies all these criteria, it is included in the lens sample,
otherwise another foreground galaxy and source pair are selected
randomly until the desired number of lens systems is obtained. For a
number of $N_{\mathrm{attempt}}$ `trial' sources, placed in an area
$A_{\mathrm{attempt}}<<A_{\mathrm{survey}}$ where
$A_{\mathrm{survey}}$ is the total area of the survey, there is a
number of $N_{\mathrm{unlensed}}$ sources that are unlensed but above
the flux limit for detection. The number of expected lenses in the
survey is then given by
$N_{\mathrm{lens}}=f_{\mathrm{lens}}N_{\mathrm{unlensed}}\times
A_{\mathrm{survey}}/A_{\mathrm{attemps}}$, with $f_{\mathrm{lens}}$
being the lensing fraction.
 
\subsection{Properties of the lens sample}

We use the procedure described above, to create a sample of
$N_{\mathrm{lens}}=400$ lens systems from our mock galaxy
catalogue. This number of lens systems was chosen to obtain a
reasonable statistical sample while keeping the time needed for
computations tractable. The total number of `trial' sources was
$N_{\mathrm{attempt}}=110000$ in a total area of
$A_{\mathrm{attempt}}=0.1384\deg^2$.  Of these `trial' sources, a
number $N_{\mathrm{unlensed}}=60\pm7$\footnote{The error bar was
  estimated by running a set of 10 trial runs, and noting the value of
  $N_{\mathrm{unlensed}}$ in each case.} were unlensed but above the detection flux
limit $F_0$. The total number of sources in the total area of
$A_{\mathrm{survey}}=651\deg^2$ that would create 400 lenses
\footnote{Here, we assume that  no sources are lensed that lie outside the area
$A_{\mathrm{attempt}}$} is $282021$, yielding a fraction of
$f_{\mathrm{lens}}=1.4\pm0.18\times10^{-3}$ of all `observed' sources
as lensed. This fraction is in good agreement with the lens fraction
observed in CLASS.

\subsubsection{Lens velocity dispersions and morphologies}

We determine the average velocity dispersion of the $400$ mock lens
systems to be
$\average{\sigma}=164.2\pm2.6\,\mathrm{km\,s^{-1}}$. This value is
consistent with but slightly lower (by about $5\,\mathrm{km\,s^{-1}}$)
than what is observed in the CLASS survey.

\begin{figure}
\epsfig{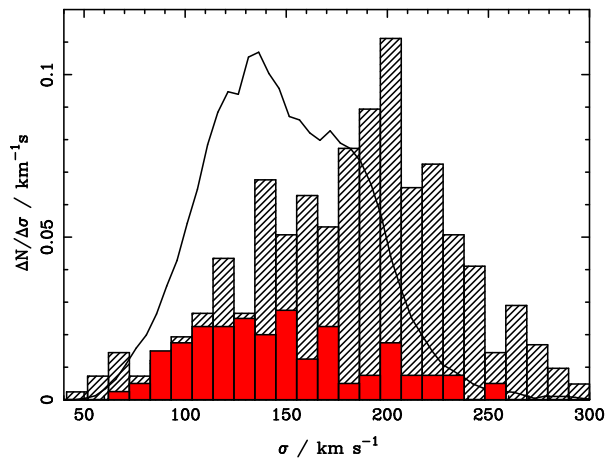}
\caption{The distribution of lens velocity dispersion in a mock lens
sample selected in the radio. The black, hatched histogram shows the velocity
dispersion distribution of all lens systems with an average velocity
dispersion of $\average{\sigma}=164.2\pm2.6\,\mathrm{km\,s^{-1}}$,
with the majority of lens galaxies ($\sim80\%$) being ellipticals. The
red, filled histogram shows the statistics for 4 image lenses only. The
solid, thin black line shows the distribution of velocity dispersions
of all ellipticals in our mock 2dF catalogue.}
\label{classvfig}
\end{figure}

The distribution of lens velocity dispersions for 2dF lenses expected
in a radio selected survey is shown in Fig.\,\ref{classvfig}. The
distribution for all systems (black) and for 4-image systems only
(red) is shown. We also show the location of lens galaxies in the
sample in Fig.\,\ref{rdistfig} in \S\,\ref{bulge} (black dots). Even
though the lenses lie on the fundamental plane, the complete sample of
lens systems has a significantly higher velocity dispersion than the
elliptical galaxy population as a whole. However, the velocity
dispersion function for 4-image lenses follows the distribution of the
galaxy velocity dispersion for all galaxies much more closely. In the
complete sample, a total fraction of about $80\pm5\%$ of lenses are
ellipticals with a bulge-to-total light ratio of $>0.6$. This fraction
decreases to $70\pm10\%$ for the 4-image system subsample. The fact
that these two fractions are very similar suggests that magnification
bias is stronger for early-type massive galaxies with a moderately 
asymmetric  potential, rather than less massive late-type systems with
strong ellipticity in the potential due to a massive disc.  

\subsubsection{Image separations and image geometries}

Using the above selection criteria, we find that the average image
separation is $\average{\Delta\theta_{\mathrm{max}}}=3.3\pm
0.2\arcsec$. Taking into account the expected difference in redshift
distributions of the lenses, $\average{z}\sim0.1$ and
$\average{z_{\mathrm{CLASS}}}\sim0.4$, and noting that the image
separations scale as $\Delta\theta_{\mathrm{max}}\propto D_{\mathrm{LS}}/D_{\mathrm{OS}}$
this compares well with the observed image separations in the CLASS
sample $\Delta\theta_{\mathrm{CLASS}}=1.13\arcsec$.

\begin{figure}
\epsfig{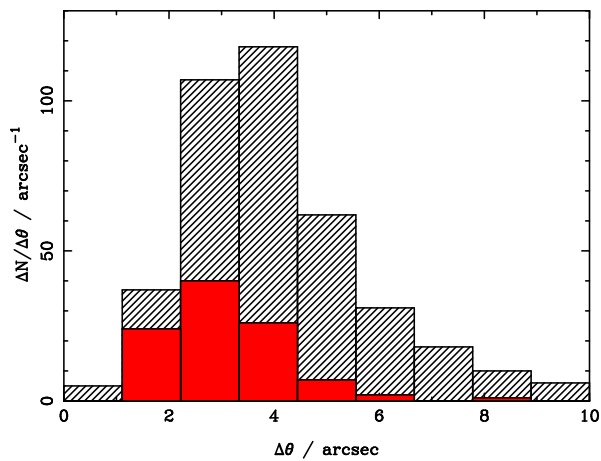}
\caption{The expected distribution of image separations in a radio
  selected survey for 2dF lenses. As in the previous figure, the black
  hatched area shows the histogram for all lens systems. The red, solid region
  shows the statistics of 4 image systems only. In contrast to
  Fig.\,\ref{erdist.fig} in \S\,\ref{maps}, this calculation includes magnification bias.}
\label{classsepsfig}
\end{figure}

The distribution of image separations produced by 2dF lenses expected
in a radio selected survey is shown in Fig.\,\ref{classsepsfig}. We
also show the image separation statistic for a subsample consisting of
4-image systems. The average maximum image separation decreases to
$\average{\Delta\theta}=2.57\pm0.6\arcsec$ when only 4-image systems
are considered. Our model for lenses in the 2dF predicts a fraction of
$70\pm5\%$ of double image systems, with a fraction of $\sim30\pm5\%$
being lens systems with four
images. Depending
on the particular choice of known lens systems to compare with, these
numbers are consistent/inconsistent with observations. The sample used
by \citet{rusin2001} contains seven quadruples and five doubles,
whereas the sample used by \citet{chae2003} and \citet{chae2005}
contain ten lenses with four quadruple image systems. These samples
differ in their selection criteria, and since the number of known lens
systems is low it is difficult to draw any firm conclusions. From our
analysis, it does appear that the number of observed quadruple systems
is slightly higher than what would be expected, even if we compare
with the sample of \citet{chae2005}. Since our model includes
ellipticities and disc components it is difficult to explain any
further asymmetry in the lensing potential needed to increase the
fraction of quadruples with anything other than the presence of other
mass concentrations in the vicinity of the primary lens. These
environmental effects are discussed in more detail in
\citet{moller2006}.

\section{Discussion and conclusions}
\label{discuss}

Surveys, like the 2dF and the SDSS are likely to contain a large
number of gravitational lens systems. We predict the number of lenses
and lensing statistics expected in the 2dF survey. The number of 
observed lensing systems is a function of the statistical properties
of the foreground galaxy population, the source population, selection
biases and cosmological parameters. Taking the cosmological parameters
as known we calculate the expected lensing statistics for point
sources at a fixed redshift using realistic multi-component galaxy
models.  Our fiducial galaxy consists of an NFW halo, an elliptical
bulge and an inclined disc. Detailed galaxy properties are predicted
using semi-analytic modeling of galaxies in the Millennium Run,
currently the largest available N-body simulation. Using ray-tracing
techniques in combination with the galaxy model we {\it predict} the
statistical lensing properties of large surveys with realistic galaxy
lens models.

We calculate the lensing cross section for a complete sample of 70,000
galaxies at redshifts of $z_{\mathrm{lens}}<0.2$ that fulfill the
selection criteria of the 2dF survey. With background sources at a
fixed redshift of $z_{\mathrm{source}}=1$ we demonstrate that the main
contribution to the lensing cross section comes from the baryonic
component of luminous, early-type galaxies. It is important to note
here that this does not mean that the lensing cross section is
insensitive to the DM component in these galaxies. In fact the
distribution of galaxy masses at a fixed luminosity,
$n(M_{\mathrm{vir}}|L)$ for a lens sample is different from that for
the complete population of early-type galaxies. As shown in
Fig.\,\ref{lumcross.fig} -- the mass function of lenses peaks at a
value that is a factor of about 2 higher than the overall galaxy
sample.  If magnification bias is not included, we find that, for
systems with maximal image separations between $1-3\arcsec$, the disc
component of spirals increases the lensing cross section by a factor of
$\sim10$. This shows that disc components of spirals are an important
contribution to statistical lensing calculations. However, when we do
include magnification bias we found that this effect is weaker due to
a trade-off between increase in cross section for late-type galaxies
on the one hand, and a much stronger magnification bias for four image
geometries in moderately elliptical massive galaxies on the other
(cf. discussion in \S\,3.3 and \S\,4.1.1).

To make firm predictions with clear selection criteria we simulate a
sample of 400 radio selected lens systems that fulfill the CLASS
selection criteria in terms of flux and magnification ratios.  We find
that, in surveys like the 2dF or SDSS, including magnification bias,
about $1.4\times10^{-3}$ of all background radio sources are lensed by
galaxies within the survey.  This number is fairly independent of the
inclusion of a disc component, but does depend crucially on the
presence of the baryonic bulge, and even more strongly, on the
magnification bias. The number of lenses per observed source increases
from $1.1\times10^{-5}$ without magnification bias to
$1.4\times10^{-3}$ including the effect of magnification bias.  Using
our model we also calculate the relative incidence of 4 and 2 image
systems. The inferred fraction of $30\pm5\%$ of quadruple systems is
consistent with observed values.

The key result of our work is that samples of lens galaxies expected in
surveys like the 2dF and the SDSS distinguish themselves from the
overall galaxy population in several important ways:
\begin{itemize}
\item A fraction of $\sim80\%$ of lenses are expected to be of
  early-type morphologies, compared to only $\sim35\%$ in the overall
  galaxy sample. This number is reduced slightly if only four image
  systems are considered.
\item The average velocity dispersion of lensing galaxies is
  $\average{\sigma}=165\,\mathrm{km\,s^{-1}}$ as opposed to
  $150\,\mathrm{km\,s^{-1}}$ for all early-type
  galaxies. Furthermore, the distribution is dramatically different --
  the distribution for lens galaxies has a maximum at $\sigma\sim200\,\mathrm{km\,s^{-1}}$,
  with a long tail towards lower velocity dispersions.
\item The peak of the luminosity distribution of lens galaxies is
  a factor of $\sim2$ higher than that of the total galaxy population. 
\item For a given luminosity, lens galaxies reside in haloes that are
  more massive by a factor of $\sim2$ on average than the overall sample of ellipticals.
\end{itemize}
Interestingly, we also find that the properties of the lens sample
differ depending on whether 4 or 2 image systems are
considered. Initially surprising is the result that 4-image systems
tend to have lower velocity dispersions; they trace the velocity
dispersion distribution of the overall early-type galaxy population
much better. This can be explained by noting that the asymmetries in
the potential have a larger effect for lower mass galaxies -- the
ratio of 4: 2 image systems is therefore expected to be much higher
for galaxies with lower velocity dispersions. In addition, the
predominant cause for 4-image lens geometries is a modest ellipticity
in the potential, which is more likely to be created by moderately
inclined discs, and hence by later-type galaxies which tend to have
lower velocity dispersions. Our assumption that asymmetries in the
potential are due to the baryonic components and that the dark matter haloes
are spherical may also affect this result. However, since lens
galaxies in the catalogue are dominated strongly by the baryonic
component, this should not change the statistics significantly.

All these differences between lens samples and the underlying galaxy
population are very important to understand if lensing studies are to
be used to obtain a better understanding of the galaxy evolution. This
holds both for statistical studies but also for studies of individual
systems. For example, single cases of lenses for which the mass
distribution is steeper than the average profile predicted by
simulations at the same mass do not point to any inconsistency; it is
absolutely crucial that the full predicted distribution functions are
considered. Our statistical analysis demonstrates clearly that {\bf
lensing galaxies are a biased population}. They preferentially sample
the high mass end of all galaxies. This systematic, in fact, limits
the unbiased study of galaxy evolution using a sample of lensing
galaxies \citep{mitchell2005,ofek2003}.

In our analysis we do not examine the effect of the environment. In
a recent study by \citet{oguri2006} it has been shown that the
presence of other mass concentrations in the vicinity can increase the
lensing cross sections for high image separations by several tens of
percent. Interestingly, this result was also previously found by
\citet{turner1984} who used a very simple mass-sheet model to
calculate the effect of environment. We address the effect of
environment, and in particular the incidence and statistical
properties of lens galaxies in groups in a subsequent publication. It
is worthwhile to note here, that for low redshifts, the environment is
expected to have a much smaller effect, as the ratio between expected
image separations of galaxy lenses and the typical distance to nearby
galaxies is much smaller at lower redshifts (at $z\sim0.05$ a distance of $1\arcsec$ corresponds to about 1\,kpc, at
$z\sim0.5$ to about 6\,kpc).

Our calculations show that selection effects, and in particular
magnification biases, have a strong effect on lens statistics. Given
that magnification bias is important, it does appear that the
distribution of the sources may also strongly influence lensing
statistics. For radio selected samples the selection criteria can be
defined clearly and the source luminosity function is known to an
adequate degree to predict lensing statistics reasonably well. In the
optical or infrared, however, both selection criteria and source
properties are not as well understood. To use lenses selected at these
wavelengths in a statistical way requires careful analysis to
disentangle the various effects, like source luminosity functions,
magnification bias and source redshift distributions.  Simulating real
imaging data in various wave-bands will be needed to test the
particular predictions of a given model of the lens and source
population.

The basis for this study has been to use one of the best models of
galaxy formation currently available and test the predicted lensing
properties of such a sample. Our choice of a semi-analytic model was
motivated strongly by the advantage that the relation between luminous
and the DM is predicted in a way that is self-consistent, easy to
interpret and use.  It is important to point out, though, that these
models do have limitations.  Even though the predictions at low
redshift appear to be consistent with observations, it is not clear if
current semi-analytic models predict all the observed properties of
galaxies at higher redshifts correctly. It is encouraging for
semi-analytic models that our results are consistent with the current
known statistical lens sample, but we add the caveat that the known
lens sample is small. Testing models of galaxy formation and evolution
using strong lensing statistics seems a promising approach; future
lens surveys will increase the number of lens systems known by a large
number and if the selection criteria in these surveys can be
understood, much can be learnt about the relation between mass and
light on galaxy scales.

\section{Summary}
\label{summary}

We predict the lensing properties of galaxies in low-redshift surveys
with $z\lesssim0.2$ using realistic galaxy mass models based on a recent
semi-analytic simulation of galaxies formation within the Millennium Run N-body
simulation. Using ray-tracing techniques we calculate the lensing
cross section for 2 and 4 image systems for a multi-component lens
model consisting of a DM halo, a bulge and a disc component.  Our key
results can be summarised as follows:
\begin{itemize}
\item The predicted lensing rate for radio sources in low redshift
  surveys like the 2dF is $1.4\times10^{-3}$.
\item A fraction of $80\pm5\%$ of all lens galaxies are predicted to
  be ellipticals.
\item The predicted average maximum image separation is $3.3\pm0.2\,\mathrm{arcsec}$.
\item The average velocity dispersion of lenses is
  $\average\sigma=164\pm2.6\,\mathrm{km\,s^{-1}}$.
\item The baryonic component is the most important contributor to the
  {\bf lensing cross sections} for galaxy scale masses, but the dark matter
  distribution also affects the {\bf lensing statistics} of composite lens
  models significantly.  
\item Lens galaxies are more luminous and, for a fixed luminosity,
  reside in dark matter haloes that have masses a factor of 2 higher than the
  overall galaxy population.
\item The velocity dispersion distribution of lens galaxies is shifted
  to significantly higher values of the velocity dispersion with respect 
  to that of a complete sample of galaxies.
\item Four image systems have, on average, a lower velocity
  dispersion, a later type morphology and a lower average maximum image
  separation than double image systems.
\end{itemize}
Our quantitative predictions are entirely consistent with the
statistics in the CLASS survey. We pioneer the use of detailed
semi-analytic models of galaxies to predict statistical lensing
properties. This approach provides testable predictions for the
lensing statistics and conversely, statistical lensing can provide
useful constraints on galaxy formation models in the near future. In ongoing
work we address in detail the effect of environment for lensing
statistics.

\section*{Acknowledgments}

We thank Simon White, Leon Koopmans, Jeremy Blaizot and Vince Eke for
many stimulating discussions during various stages of this project. We
are also grateful to Simon White and Vince Eke for useful comments on
the manuscript. We also thank Volker Springel and the Virgo Consortium
for the use of the Millennium Run simulation.

\bibliographystyle{mn}
\bibliography{mn-jour,Bibs}

\begin{thebibliography}{67}
\expandafter\ifx\csname natexlab\endcsname\relax\def\natexlab#1{#1}\fi

\bibitem[{{Adelman-McCarthy} {et~al.}(2006){Adelman-McCarthy}, {Ag{\"u}eros},
  {Allam}, {Anderson}, {Anderson}, {Annis}, {Bahcall}, {Baldry}, {Barentine},
  {Berlind}, {Bernardi}, {Blanton}, {Boroski}, {Brewington}, {Brinchmann},
  {Brinkmann}, {Brunner}, {Budav{\'a}ri}, {Carey}, {Carr}, {Castander},
  {Connolly}, {Csabai}, {Czarapata}, {Dalcanton}, {Doi}, {Dong}, {Eisenstein},
  {Evans}, {Fan}, {Finkbeiner}, {Friedman}, {Frieman}, {Fukugita}, {Gillespie},
  {Glazebrook}, {Gray}, {Grebel}, {Gunn}, {Gurbani}, {de Haas}, {Hall},
  {Harris}, {Harvanek}, {Hawley}, {Hayes}, {Hendry}, {Hennessy}, {Hindsley},
  {Hirata}, {Hogan}, {Hogg}, {Holmgren}, {Holtzman}, {Ichikawa}, {Ivezi{\'c}},
  {Jester}, {Johnston}, {Jorgensen}, {Juri{\'c}}, {Kent}, {Kleinman}, {Knapp},
  {Kniazev}, {Kron}, {Krzesinski}, {Kuropatkin}, {Lamb}, {Lampeitl}, {Lee},
  {Leger}, {Lin}, {Long}, {Loveday}, {Lupton}, {Margon},
  {Mart{\'{\i}}nez-Delgado}, {Mandelbaum}, {Matsubara}, {McGehee}, {McKay},
  {Meiksin}, {Munn}, {Nakajima}, {Nash}, {Neilsen}, {Newberg}, {Newman},
  {Nichol}, {Nicinski}, {Nieto-Santisteban}, {Nitta}, {O'Mullane}, {Okamura},
  {Owen}, {Padmanabhan}, {Pauls}, {Peoples}, {Pier}, {Pope}, {Pourbaix},
  {Quinn}, {Richards}, {Richmond}, {Rockosi}, {Schlegel}, {Schneider},
  {Schroeder}, {Scranton}, {Seljak}, {Sheldon}, {Shimasaku}, {Smith}, {Smol{\v
  c}i{\'c}}, {Snedden}, {Stoughton}, {Strauss}, {SubbaRao}, {Szalay},
  {Szapudi}, {Szkody}, {Tegmark}, {Thakar}, {Tucker}, {Uomoto}, {Vanden Berk},
  {Vandenberg}, {Vogeley}, {Voges}, {Vogt}, {Walkowicz}, {Weinberg}, {West},
  {White}, {Xu}, {Yanny}, {Yocum}, {York}, {Zehavi}, {Zibetti}, \&
  {Zucker}}]{abdel2006}
{Adelman-McCarthy} J.~K., et al., 2006, \apjs, 162, 38

\bibitem[{{Bartelmann} {et~al.}(1998){Bartelmann}, {Huss}, {Colberg},
  {Jenkins}, \& {Pearce}}]{bartelmann1998}
{Bartelmann} M., {Huss} A., {Colberg} J.~M., {Jenkins} A., {Pearce} F.~R.,
  1998, \aap, 330, 1

\bibitem[{{Bernardi} {et~al.}(2003{\natexlab{a}}){Bernardi}, {Sheth}, {Annis},
  {Burles}, {Eisenstein}, {Finkbeiner}, {Hogg}, {Lupton}, {Schlegel},
  {SubbaRao}, {Bahcall}, {Blakeslee}, {Brinkmann}, {Castander}, {Connolly},
  {Csabai}, {Doi}, {Fukugita}, {Frieman}, {Heckman}, {Hennessy}, {Ivezi{\'c}},
  {Knapp}, {Lamb}, {McKay}, {Munn}, {Nichol}, {Okamura}, {Schneider}, {Thakar},
  \& {York}}]{bernardi2003c}
{Bernardi} M., et al., 2003{\natexlab{a}}, \aj,
  125, 1866

\bibitem[{{Bernardi} {et~al.}(2003{\natexlab{b}}){Bernardi}, {Sheth}, {Annis},
  {Burles}, {Eisenstein}, {Finkbeiner}, {Hogg}, {Lupton}, {Schlegel},
  {SubbaRao}, {Bahcall}, {Blakeslee}, {Brinkmann}, {Castander}, {Connolly},
  {Csabai}, {Doi}, {Fukugita}, {Frieman}, {Heckman}, {Hennessy}, {Ivezi{\'c}},
  {Knapp}, {Lamb}, {McKay}, {Munn}, {Nichol}, {Okamura}, {Schneider}, {Thakar},
  \& {York}}]{bernardi2003a}
{Bernardi} M., et al., 2003{\natexlab{b}}, \aj, 125, 1817

\bibitem[{{Blain}(1998)}]{blain1998}
{Blain} A.~W., 1998, \mnras, 295, 92

\bibitem[{{Bower} {et~al.}(2006){Bower}, {Benson}, {Malbon}, {Helly}, {Frenk},
  {Baugh}, {Cole}, \& {Lacey}}]{bower2006}
{Bower} R.~G., {Benson} A.~J., {Malbon} R., {Helly} J.~C., {Frenk} C.~S.,
  {Baugh} C.~M., {Cole} S., {Lacey} C.~G., 2006, \mnras, 659

\bibitem[{{Broadhurst} {et~al.}(2005){Broadhurst}, {Ben{\'{\i}}tez}, {Coe},
  {Sharon}, {Zekser}, {White}, {Ford}, {Bouwens}, {Blakeslee}, \&
  {Clampin}}]{broadhurst2005}
{Broadhurst} T., {Ben{\'{\i}}tez} N., {Coe} D., {Sharon} K., {Zekser} K.,
  {White} R., {Ford} H., {Bouwens} R., {Blakeslee} J., {Clampin}, 2005, \apj,
  621, 53

\bibitem[{{Bullock} {et~al.}(2001){Bullock}, {Kolatt}, {Sigad}, {Somerville},
  {Kravtsov}, {Klypin}, {Primack}, \& {Dekel}}]{bullock2001}
{Bullock} J.~S., {Kolatt} T.~S., {Sigad} Y., {Somerville} R.~S., {Kravtsov}
  A.~V., {Klypin} A.~A., {Primack} J.~R., {Dekel} A., 2001, \mnras, 321, 559

\bibitem[{{Cappellari} {et~al.}(2006){Cappellari}, {Bacon}, {Bureau}, {Damen},
  {Davies}, {de Zeeuw}, {Emsellem}, {Falc{\'o}n-Barroso}, {Krajnovi{\'c}},
  {Kuntschner}, {McDermid}, {Peletier}, {Sarzi}, {van den Bosch}, \& {van de
  Ven}}]{cappellari2006}
{Cappellari} et al., 2006, \mnras, 366, 1126

\bibitem[{{Chae}(2003)}]{chae2003}
{Chae} K.-H., 2003, \mnras, 346, 746

\bibitem[{{Chae}(2005)}]{chae2005}
{Cahe} K.-H., 2005, \apj, 630, 764

\bibitem[{{Cohn} {et~al.}(2001){Cohn}, {Kochanek}, {McLeod}, \&
  {Keeton}}]{cohn2001}
{Cohn} J.~D., {Kochanek} C.~S., {McLeod} B.~A., {Keeton} C.~R., 2001, \apj,
  554, 1216

\bibitem[{{Colless} {et~al.}(2001){Colless}, {Dalton}, {Maddox}, {Sutherland},
  {Norberg}, {Cole}, {Bland-Hawthorn}, {Bridges}, {Cannon}, {Collins}, {Couch},
  {Cross}, {Deeley}, {De Propris}, {Driver}, {Efstathiou}, {Ellis}, {Frenk},
  {Glazebrook}, {Jackson}, {Lahav}, {Lewis}, {Lumsden}, {Madgwick}, {Peacock},
  {Peterson}, {Price}, {Seaborne}, \& {Taylor}}]{colless2001}
{Colless} M., et al., 2001, \mnras, 328, 1039

\bibitem[{{Cooray} \& {Cen}(2005)}]{cooray2005}
{Cooray} A., {Cen} R., 2005, \apjl, 633, L69

\bibitem[{{Croton} {et~al.}(2006){Croton}, {Springel}, {White}, {De Lucia},
  {Frenk}, {Gao}, {Jenkins}, {Kauffmann}, {Navarro}, \& {Yoshida}}]{croton2006}
{Croton} et al., 2006,
  \mnras, 365, 11

\bibitem[{{Dalal} {et~al.}(2005){Dalal}, {Hennawi}, \& {Bode}}]{dalal2005}
{Dalal} N., {Hennawi} J., {Bode} P., 2005, \apj, 622, 99 

\bibitem[{{de Jong} {et~al.}(2004){de Jong}, {Simard}, {Davies}, {Saglia},
  {Burstein}, {Colless}, {McMahan}, \& {Wegner}}]{dejong2004}
{de Jong} R.~S., {Simard} L., {Davies} R.~L., {Saglia} R.~P., {Burstein} D.,
  {Colless} M., {McMahan} R., {Wegner} G., 2004, \mnras, 355, 1155

\bibitem[{{de Vaucouleurs}(1948)}]{devaucouleurs1948}
{de Vaucouleurs} G., 1948, Annales d'Astrophysique, 11, 247

\bibitem[{{Eke} {et~al.}(2001){Eke}, {Navarro}, \& {Steinmetz}}]{eke2001}
{Eke} V.~R., {Navarro} J.~F., {Steinmetz} M., 2001, \apj, 554, 114

\bibitem[{{Ferreras} {et~al.}(2005){Ferreras}, {Saha}, \&
  {Williams}}]{ferreras2005}
{Ferreras} I., {Saha} P., {Williams} L.~L.~R., 2005, \apjl, 623, L5

\bibitem[{{Fukugita} {et~al.}(1992){Fukugita}, {Futamase}, {Kasai}, \&
  {Turner}}]{fukugita1992}
{Fukugita} M., {Futamase} T., {Kasai} M., {Turner} E.~L., 1992, \apj, 393, 3

\bibitem[{{Gerhard} {et~al.}(2001){Gerhard}, {Kronawitter}, {Saglia}, \&
  {Bender}}]{gerhard2001}
{Gerhard} O., {Kronawitter} A., {Saglia} R.~P., {Bender} R., 2001, \aj, 121,
  1936

\bibitem[{{Helbig} {et~al.}(1999){Helbig}, {Marlow}, {Quast}, {Wilkinson},
  {Browne}, \& {Koopmans}}]{helbig1999}
{Helbig} P., {Marlow} D., {Quast} R., {Wilkinson} P.~N., {Browne} I. W.~A.,
  {Koopmans} L. V.~E., 1999, \aaps, 136, 297

\bibitem[{{Huterer} {et~al.}(2005){Huterer}, {Keeton}, \& {Ma}}]{huterer2005}
{Huterer} D., {Keeton} C.~R., {Ma} C.-P., 2005, \apj, 624, 34

\bibitem[{{Jorgensen} \& {Franx}(1994)}]{jorgensen1994}
{Jorgensen} I., {Franx} M., 1994, \apj, 433, 553

\bibitem[{{Kauffmann} {et~al.}(1993){Kauffmann}, {White}, \&
  {Guiderdoni}}]{kauffmann1993}
{Kauffmann} G., {White} S.~D.~M., {Guiderdoni} B., 1993, \mnras, 264, 201

\bibitem[{{Kauffmann} {et~al.}(2004){Kauffmann}, {White}, {Heckman},
  {M{\'e}nard}, {Brinchmann}, {Charlot}, {Tremonti}, \&
  {Brinkmann}}]{kauffmann2004}
{Kauffmann} G., {White} S.~D.~M., {Heckman} T.~M., {M{\'e}nard} B.,
  {Brinchmann} J., {Charlot} S., {Tremonti} C., {Brinkmann} J., 2004, \mnras,
  353, 713

\bibitem[{{Keeton} \& {Zabludoff}(2004)}]{keeton2004}
{Keeton} C.~R., {Zabludoff} A.~I., 2004, \apj, 612, 660

\bibitem[{{King} \& {Browne}(1996)}]{king1996}
{King} L.~J., {Browne} I.~W.~A., 1996, \mnras, 282, 67

\bibitem[{{Kneib} {et~al.}(2003){Kneib}, {Hudelot}, {Ellis}, {Treu}, {Smith},
  {Marshall}, {Czoske}, {Smail}, \& {Natarajan}}]{kneib2003}
{Kneib} J.-P., {Hudelot} P., {Ellis} R.~S., {Treu} T., {Smith} G.~P.,
  {Marshall} P., {Czoske} O., {Smail} I., {Natarajan} P., 2003, \apj, 598, 804

\bibitem[{{Kochanek}(1993)}]{kochanek1993}
{Kochanek} C.~S., 1993, \mnras, 261, 453

\bibitem[{{Kochanek}(1996)}]{kochanek1996}
{Kochanek} C.~S., 1996, \apj, 473, 595

\bibitem[{{Koopmans} \& {Treu}(2002)}]{koopmans2002a}
{Koopmans} L.~.~V.~E., {Treu} T., 2002, \apjl, 568, L5

\bibitem[{{Koopmans} \& {Treu}(2003)}]{koopmans2003}
{Koopmans} L.~V.~E., {Treu} T., 2003, \apj, 583, 606

\bibitem[{{Kundic} {et~al.}(1997{\natexlab{a}}){Kundic}, {Hogg}, {Blandford},
  {Cohen}, {Lubin}, \& {Larkin}}]{kundic1997b}
{Kundic} T., {Hogg} D.~W., {Blandford} R.~D., {Cohen} J.~G., {Lubin} L.~M.,
  {Larkin} J.~E., 1997{\natexlab{a}}, \aj, 114, 2276

\bibitem[{{Kundic} {et~al.}(1997{\natexlab{b}}){Kundic}, {Turner}, {Colley},
  {Gott}, {Rhoads}, {Wang}, {Bergeron}, {Gloria}, {Long}, {Malhotra}, \&
  {Wambsganss}}]{kundic1997a}
{Kundic} T., et al., 1997{\natexlab{b}}, \apj, 482, 75

\bibitem[{{Maoz} {et~al.}(1993){Maoz}, {Bahcall}, {Schneider}, {Bahcall},
  {Djorgovski}, {Doxsey}, {Gould}, {Kirhakos}, {Meylan}, \& {Yanny}}]{maoz1993}
{Maoz} D., et al. , 1993,
  \apj, 409, 28

\bibitem[{{Meneghetti} {et~al.}(2005){Meneghetti}, {Jain}, {Bartelmann}, \&
  {Dolag}}]{meneghetti2005}
{Meneghetti} M., {Jain} B., {Bartelmann} M., {Dolag} K., 2005, MNRAS,
  362, 1301

\bibitem[{{Mitchell} {et~al.}(2005){Mitchell}, {Keeton}, {Frieman}, \&
  {Sheth}}]{mitchell2005}
{Mitchell} J.~L., {Keeton} C.~R., {Frieman} J.~A., {Sheth} R.~K., 2005, \apj,
  622, 81

\bibitem[{{M{\"o}ller} \& {Blain}(1998)}]{moller1998}
{M{\"o}ller} O., {Blain} A.~W., 1998, \mnras, 299, 845

\bibitem[{{M{\"o}ller} \& {Blain}(2001)}]{moller2001}
M{\"oller}, O., {Blain} A.~W., 2001, \mnras, 327, 339

\bibitem[{{M{\"o}ller} {et~al.}(2003){M{\"o}ller}, {Hewett}, \&
  {Blain}}]{moller2003}
{M{\"o}ller} O., {Hewett} P., {Blain} A.~W., 2003, \mnras, 345, 1

\bibitem[{{M{\"o}ller} {et~al.}(2002){M{\"o}ller}, {Natarajan}, {Kneib}, \&
  {Blain}}]{moller2002}
{M{\"o}ller} O., {Natarajan} P., {Kneib} J.-P., {Blain} A.~W., 2002, \apj, 573,
  562

\bibitem[{{M{\"o}ller}(2006)}]{moller2006}
{M{\"o}ller} O. e.~a., 2006, Lensing by groups in 2df and sdss, \mnras, in
  preparation

\bibitem[{{Navarro} {et~al.}(1997){Navarro}, {Frenk}, \& {White}}]{navarro1997}
{Navarro} J.~F., {Frenk} C.~S., {White} S. D.~M., 1997, \apj, 490, 493

\bibitem[{{Norberg} {et~al.}(2002){Norberg}, {Cole}, {Baugh}, {Frenk},
  {Baldry}, {Bland-Hawthorn}, {Bridges}, {Cannon}, {Colless}, {Collins},
  {Couch}, {Cross}, {Dalton}, {De Propris}, {Driver}, {Efstathiou}, {Ellis},
  {Glazebrook}, {Jackson}, {Lahav}, {Lewis}, {Lumsden}, {Maddox}, {Madgwick},
  {Peacock}, {Peterson}, {Sutherland}, \& {Taylor}}]{norberg2002}
{Norberg} P., et al., 2002, \mnras, 336, 907

\bibitem[{{Ofek} {et~al.}(2003){Ofek}, {Rix}, \& {Maoz}}]{ofek2003}
{Ofek} E.~O., {Rix} H.-W., {Maoz} D., 2003, \mnras, 343, 639

\bibitem[{{Oguri}(2006)}]{oguri2006}
{Oguri} M., 2006, \mnras, 227

\bibitem[{{Oguri} {et~al.}(2005){Oguri}, {Keeton}, \& {Dalal}}]{oguri2005}
{Oguri} M., {Keeton} C.~R., {Dalal} N., 2005, \mnras, 364, 1451

\bibitem[{{Pierce} \& {Tully}(1992)}]{pierce1992}
{Pierce} M.~J., {Tully} R.~B., 1992, \apj, 387, 47

\bibitem[{{Rhee}(1991)}]{rhee1991}
{Rhee} G., 1991, \nat, 350, 211

\bibitem[{{Rusin} \& {Ma}(2001)}]{rusin2001a}
{Rusin} D., {Ma} C.-P., 2001, \apjl, 549, L33

\bibitem[{{Rusin} \& {Tegmark}(2001)}]{rusin2001}
{Rusin} D., {Tegmark} M., 2001, \apj, 553, 709

\bibitem[{{Schechter} {et~al.}(1997){Schechter}, {Bailyn}, {Barr}, {Barvainis},
  {Becker}, {Bernstein}, {Blakeslee}, {Bus}, {Dressler}, {Falco}, {Fesen},
  {Fischer}, {Gebhardt}, {Harmer}, {Hewitt}, {Hjorth}, {Hurt}, {Jaunsen},
  {Mateo}, {Mehlert}, {Richstone}, {Sparke}, {Thorstensen}, {Tonry}, {Wegner},
  {Willmarth}, \& {Worthey}}]{schechter1997}
{Schechter} P.~L., et al., 1997, \apjl, 475, L85+

\bibitem[{{Somerville} \& {Primack}(1999)}]{somerville1999}
{Somerville} R.~S., {Primack} J.~R., 1999, \mnras, 310, 1087

\bibitem[{{Spergel} {et~al.}(2006){Spergel}, {Bean}, {Dore'}, {Nolta},
  {Bennett}, {Hinshaw}, {Jarosik}, {Komatsu}, {Page}, {Peiris}, {Verde},
  {Barnes}, {Halpern}, {Hill}, {Kogut}, {Limon}, {Meyer}, {Odegard}, {Tucker},
  {Weiland}, {Wollack}, \& {Wright}}]{spergel2006}
{Spergel} D.~N., et al., 2006, preprint (astro-ph/0603449)

\bibitem[{{Springel} {et~al.}(2005){Springel}, {White}, {Jenkins}, {Frenk},
  {Yoshida}, {Gao}, {Navarro}, {Thacker}, {Croton}, {Helly}, {Peacock}, {Cole},
  {Thomas}, {Couchman}, {Evrard}, {Colberg}, \& {Pearce}}]{springel2005}
{Springel} V., et al., 2005, \nat, 435, 629

\bibitem[{{Treu} \& {Koopmans}(2003)}]{treu2003}
{Treu} T., {Koopmans} L.~V.~E., 2003, \mnras, 343, L29

\bibitem[{{Tully} \& {Fisher}(1977)}]{tully1977}
{Tully} R.~B., {Fisher} J.~R., 1977, \aap, 54, 661

\bibitem[{{Tully} \& {Pierce}(2000)}]{tully2000}
{Tully} R.~B., {Pierce} M.~J., 2000, \apj, 533, 744

\bibitem[{{Turner} {et~al.}(1984){Turner}, {Ostriker}, \& {Gott}}]{turner1984}
{Turner} E.~L., {Ostriker} J.~P., {Gott} J.~R., 1984, \apj, 284, 1

\bibitem[{{Vale} \& {Ostriker}(2004)}]{vale2004}
{Vale} A., {Ostriker} J.~P., 2004, \mnras, 353, 189

\bibitem[{{Willott} {et~al.}(2001){Willott}, {Rawlings}, {Blundell}, {Lacy}, \&
  {Eales}}]{willott2001}
{Willott} C.~J., {Rawlings} S., {Blundell} K.~M., {Lacy} M., {Eales} S.~A.,
  2001, \mnras, 322, 536

\bibitem[{{Wucknitz}(2002)}]{wucknitz2002}
{Wucknitz} O., 2002, \mnras, 332, 951

\bibitem[{{Wucknitz}(2004)}]{wucknitz2004}
---, 2004, \mnras, 349, 1

\bibitem[{{Wyithe} {et~al.}(2002){Wyithe}, {Agol}, {Turner}, \&
  {Schmidt}}]{wyithe2002}
{Wyithe} J.~S.~B., {Agol} E., {Turner} E.~L., {Schmidt} R.~W., 2002, \mnras,
  330, 575

\bibitem[{{Zhao} \& {Qin}(2003)}]{zhao2003}
{Zhao} H., {Qin} B., 2003, \apj, 582, 2

\end{thebibliography}
\end{document}